\documentclass{aa}  

\usepackage{natbib}
\bibpunct{(}{)}{;}{a}{}{,} 
\usepackage{graphicx}

\usepackage{txfonts}
\usepackage{xcolor}

\begin{document}

\title{Vortices evolution in the solar atmosphere}

\subtitle{A dynamical equation for the swirling strength}

\author{José R.\,Canivete Cuissa\inst{1,}\inst{2}
	\and
	Oskar Steiner\inst{1,}\inst{3}
}

\offprints{J.\,R.~Canivete Cuissa, \email{jcanivete@ics.uzh.ch}, O.~Steiner, \email{steiner@leibniz-kis.de}}

 \institute{Istituto Ricerche Solari Locarno (IRSOL),
	Via Patocchi 57, 6605 Locarno-Monti, Switzerland
	\and
	Center for Theoretical Astrophysics and Cosmology, Institute for Computational Science (ICS),\\University of Zurich, Winterthurerstrasse 190, 8057 Z{\"u}rich, Switzerland
	\and
	Leibniz-Institut f\"ur Sonnenphysik (KIS), 
	Sch\"oneckstrasse 6, 79104 Freiburg i.Br., Germany
}
\date{Received xxx / Accepted yyy}

\abstract 
{} 
{We study vortex dynamics in the solar atmosphere by employing and deriving the analytical evolution equations of two vortex identification criteria.}
{The two criteria used are vorticity and the swirling strength. Vorticity can be biased in the presence of shear flows, but its dynamical equation is well known; the swirling strength is a more precise criterion for the identification of vortical flows, but its evolution equation is not known yet. Therefore, we explore the possibility of deriving a dynamical equation for the swirling strength. We then apply the two equations to analyze radiative magneto-hydrodynamical simulations of the solar atmosphere produced with the CO$^5$BOLD code.} 
{We present a detailed review of the swirling strength criterion and the mathematical derivation of its evolution equation. This equation did not exist in the literature before and it constitutes a novel tool that is suitable for the analysis of a wide range of problems in (magneto-)hydrodynamics. By applying this equation to numerical models, we find that hydrodynamical and magnetic baroclinicities are the driving physical processes responsible for vortex generation in the convection zone and the photosphere. Higher up in the chromosphere, the magnetic terms alone dominate.  Moreover, we find that the swirling strength is produced at small scales in a chaotic fashion, especially inside magnetic flux concentrations.}
{The swirling strength represents an appropriate criterion for the identification of vortices in turbulent flows, such as those in the solar atmosphere. Moreover, its evolution equation, which is derived in this paper, is pivotal for obtaining precise information about the dynamics of these vortices and the physical mechanisms responsible for their production and evolution. Since this equation is available, the swirling strength is now the ideal quantity to study the dynamics of vortices in (magneto-)hydrodynamics.}

\maketitle


\section{Introduction}

Vortical motions in the solar atmosphere have received increasing attention in recent years. While a single vortical photospheric flow of multi-granular spatial size was first observed by 
\citet{1988Natur.335..238B}, 
\citet{2008ApJ...687L.131B} 
and \citet{2010A&A...513L...6B} 
found small photospheric swirls of integranular scale by tracking magnetic bright points on their way to becoming engulfed by a downdraft. Subsequently, \citet{2010ApJ...723L.139B} 
and \citet{2011MNRAS.416..148V} 
detected vortical motions of intergranular scale relying on the methods of local correlation tracking (LCT). Small-scale photospheric vortical flows were also reported, for example, by 
\citet{2009A&A...493L..13A} 
and 
\citet{2011A&A...531L...9M}. 
In the chromosphere, 
\citet{2009A&A...507L...9W} 
observed narrow rotating rings and ring fragments with the CRISP Fabry P\'erot System of the Swedish Solar Telescope (SST) by recording narrow band filtergrams in the line core of \ion{Ca}{i} 854.2\,$\mathrm{nm}$ of a quite Sun region inside a coronal hole. The rings had a diameter of approximately 4\,$\mathrm{arcsec}$, corresponding to the size of a super large terrestrial hurricane.
More recently, 
\citet{2018A&A...618A..51T, 2019A&A...623A.160T} 
found a persistent chromospheric swirl with a diameter of 6\,$\mathrm{arcsec}$ and a lifetime of at least 1.7\,$\mathrm{h}$.

Subsequent observations and numerical simulations by 
\citet{2012Natur.486..505W} 
led to the conclusion that chromospheric swirls are the observational signatures of rotating coherent magnetic structures, which also bear imprints in the transition zone and corona. 
In particular, the simulations revealed that the chromospheric plasma is dragged by rotating ``frozen-in'' magnetic field structures, which have their footpoints in the intergranular lanes of the photosphere and upper convection zone 
\citep{2014PASJ...66S..10W}. 
Given that their shape resembles Earth's atmosphere phenomena, they are usually referred to as (solar) ``magnetic tornadoes'' \citep{2013JPhCS.440a2005W}. 
In contrast to that scenario,  
\citet{2013ApJ...776L...4S} 
argue that the photospheric vortex-like structures observed in numerical simulations are in reality torsional Alfvén waves propagating along the magnetic flux tubes which do not exhibit a tornado-like behavior.  

One interesting aspect of magnetic tornadoes is that, by coupling the convection zone and the photosphere to the chromosphere and corona, they could act as channels for the transport of energy into the upper layers of the solar atmosphere, thus contributing to their heating. This idea has been proposed by \citet{2012Natur.486..505W}, 
as they estimated through numerical simulations that magnetic tornadoes could carry a mean positive Poynting flux in the vertical direction of $440\,\mathrm{W\,m^{-2}}$ through torsional Alfvén waves. 

It is not yet clear how magnetic tornadoes form. One possible solution is given by purely hydrodynamical vortices, which are observable as nonmagnetic bright points in simulations \citep{2016A&A...596A..43C}, 
 which form by the conservation of angular momentum as the plasma sinks in the downdrafts of photospheric intergranular lanes. This ``bathtub'' effect would capture magnetic fields that are frozen in the accreting
plasma, making it rotate 
\citep{2012Natur.486..505W, 2014PASJ...66S..10W}. 

One way of investigating the true nature of vortex flows is to study their dynamical generation in numerical simulations. \citet{1998ApJ...499..914S} 
have shown with magnetic field-free hydrodynamical simulations that the generation of vorticity in the upper convection zone is principally due to baroclinicity, while 
\citet{2011A&A...526A...5S,2012ASPC..463..107S} 
used radiative magneto-hydrodynamical (MHD) simulations to demonstrate that intergranular photospheric vortices are more efficiently produced by magnetic tension effects. All of these studies rely on vorticity as the criterion to identify the swirling motions in the simulations and on the vorticity equation to study their dynamics. However, it is well known that vorticity is not the optimal quantity to identify swirls in turbulent flows as it cannot differentiate between actual vortices and shear flows. This could lead to a misinterpretation of the numerical simulations and therefore to wrong conclusions. 

An unambiguous vortex identification criterion is still an open problem in fluid mechanics 
\citep[see, e.g.,][]{Kolar2007,2019JHyDy..31..205L}, 
and numerous methods have been proposed and used. A promising candidate for solar atmospheric applications is the ``swirling strength'' \citep{1999JFM...387..353Z}. 
This quantity, which is based on the eigenanalysis of the velocity gradient tensor, does not detect shear flows, is applicable to compressible hydrodynamics 
\citep{2009AIAAJ..47..473K}, 
and has already been used in the context of solar vortices by \citet{2011A&A...533A.126M} 
and \cite{2017A&A...601A.135K}. 
Nevertheless, a necessary ingredient for the study of the dynamics of vortical flows is the possibility to derive an analytical evolution equation for the criteria one seeks to use. The evolution equation for the vorticity is well known, easy to derive, and it is in fact used in the literature. Instead, there is no reference for an equation describing the time evolution of the swirling strength. Such an equation is necessary to understand whether the results obtained with the vorticity equation are biased by the ubiquitous shear flows in the solar atmosphere, or not.  

In this paper, we review in detail the properties of the swirling strength criterion and we derive a new analytical equation for its evolution, the ``swirling equation''. This equation is a new tool in the general context of vortices in (M)HD, which we apply to the particular case of numerical simulations of the solar atmosphere. To provide more details, we compare the vortex generation given by the vorticity equation and the swirling equation at different heights in the atmosphere, finding substantial differences, in particular in the upper convection zone. Moreover, we show a typical configuration of each one of the terms appearing in the swirling equation in the vicinity of a magnetic flux concentration, in order to gain insights into the physical processes at the origin of the vortices in these regions. 

The paper is organized as follows: in Sect.\,\ref{sec:vortexcriteria} we present and discuss the two criteria, vorticity and swirling strength, while in Sect.\,\ref{sec:swirlingequation} we derive in detail the swirling equation. In Sect.\,\ref{sec:results} we firstly study two simple toy-models in order to better understand the advantages of the swirling strength, then we use its evolution equation as a tool to analyze vortex generation in the numerical simulations. Finally, we summarize and conclude in Sect.\,\ref{sec:conclusion}.


\section{Vortex identification criteria}\label{sec:vortexcriteria}

In order to study the evolution of small-scale vortices in the solar atmosphere, we need a dynamical description of the criteria chosen for their identification. For this purpose, we start from the momentum equation of the ideal MHD system, 
\begin{equation}
  \frac{\rm d}{{\rm d}t}\vec{v} = -\frac{1}{\rho}\nabla(p_{\rm g} + p_{\rm m}) + \frac{1}{\rho}(\vec{B}\cdot\nabla)\vec{B} - \nabla \Phi\,, \label{eq:MHD_momentum_eq}
\end{equation}
where $\vec{v}$ is the velocity vector, $\rho$ is the plasma density, $\vec{B}$ is the magnetic field\footnote{We note that the magnetic field is normalized by a factor $\sqrt{(4\pi)}$ to lighten up the notation.} and $\Phi$ is a potential of conservative forces. We also have two pressure terms: the first one, $p_{\rm g}$, is the usual atmospheric plasma pressure, while the latter represents the pressure due to the magnetic field, $p_{\rm m} = \vec{B}^2/2$. 
Finally, ${\rm d}/{\rm d}t$ is the total material derivative, which is related to the partial time derivative $\partial_t$ through
\begin{equation}
  \frac{\rm d}{{\rm d}t} = \partial_t + \vec{v}\cdot\nabla\,. 
  \label{eq:mat_deriv}
\end{equation}
From Eq.\,(\ref{eq:MHD_momentum_eq}) we can derive dynamical equations of quantities which are sensible to the presence of rotational motions and can therefore be used as criteria for vortex identification. 

Despite the very intuitive idea of vortex, its mathematical definition and, consequently, the choice of adequate criteria, reveals to be a complicated task, which is still under debate in fluid mechanics researches. Multiple criteria have been proposed in the past and new ones are yet to come. Between the most used in solar physics, we mention here the $\lambda_2$-criterion 
\citep{1995JFM...285...69J}, 
the $\Gamma$ detection functions 
\citep{2001MeScT..12.1422G}, 
and the Lagrangian Averaged Vorticity Deviation (LAVD) technique \citep{2016JFM...795..136H}. 
However, for the purposes of this paper, we are only interested in two methods: the well-known vorticity $\vec{\omega}$ and the swirling strength $\lambda$
\citep{1999JFM...387..353Z}. 


\subsection{Vorticity}
Vorticity is possibly the most common criterion in vortex identification. It is intuitively defined as the curl of the velocity field, $\vec{\omega} \coloneqq \nabla\times\vec{v}$. The direction of the vector indicates the rotation axis and the orientation of the vortex through the right-hand rule. Moreover, the module is directly linked to the rotation period of a rigid body, $T = 4\pi/|\vec{\omega}|$. An interesting feature of vorticity is its evolution equation, which can be derived by taking the curl of Eq.\,(\ref{eq:MHD_momentum_eq}),  
\setlength{\arraycolsep}{3pt}
\begin{eqnarray}	
  \frac{\rm d}{{\rm d}t}\vec{\omega} &=&     
  -\overbrace{\vphantom{\frac{1}{2}}
  \vec{\omega}(\nabla\cdot\vec{v})}^{{\textstyle T_{\rm \vec{\omega}}^0}} 
  +\overbrace{\vphantom{\frac{1}{2}}
  (\vec{\omega}\cdot\nabla)\vec{v}}^{{\textstyle T_{\rm \vec{\omega}}^1}} 
  -\overbrace{\nabla\frac{1}{\rho}\times
  \nabla p_{\rm g}}^{{\textstyle T_{\rm \vec{\omega}}^2}}
  \nonumber\\
  &&
  -\underbrace{\nabla\frac{1}{\rho}\times\left[\nabla p_{\rm m} 
  -(\vec{B}\cdot\nabla)\vec{B}\right]}_{{\textstyle T_{\rm \vec{\omega}}^3}} 
  +\underbrace{\frac{1}{\rho}\nabla\times
  (\vec{B}\cdot\nabla)\vec{B}}_{{\textstyle T_{\rm \vec{\omega}}^4}}\,, \label{eq:vorticity}
\end{eqnarray}
and which is usually called the ``vorticity equation''. 

From Eq.\,(\ref{eq:vorticity}), one can identify the physical mechanisms related to the generation or destruction of vorticity in MHD. Following 
\citet{2011A&A...526A...5S}, 
we label different terms of the vorticity equation according to their physical meaning.
The first two terms, $T^0_{\rm \vec{\omega}}$ and $T^1_{\rm \vec{\omega}}$, are directly related to vorticity and are called respectively vortex stretching and vortex tilting terms. We would like to point out that $T^0_{\rm \vec{\omega}}$ is not present in \citet{2011A&A...526A...5S} 
since we adopt a different prescription for the left-hand side of the equation. The term $T^2_{\rm \vec{\omega}}$ is the hydrodynamic baroclinic term, which is responsible for the generation of vorticity when the gradients of density and pressure are not parallel. Finally, we have two magnetic terms, $T^3_{\rm \vec{\omega}}$ and $T^4_{\rm \vec{\omega}}$. The first is the magnetic equivalent of $T^2_{\rm \vec{\omega}}$ and shall be denoted as magnetic baroclinic term, while the latter is related to magnetic tension. 

This approach is used in 
\citet{2011A&A...526A...5S,2012ASPC..463..107S} 
to study the physical nature of magnetic photospheric intergranular vortices. However, we argue that these results could be biased by the fact that vorticity is not a reliable criterion for the identification of swirls in turbulent flows. It is in fact known that vorticity can assume large values also in the presence of shear flows \citep[see e.g.,][]{1995JFM...285...69J}, 
which can be ubiquitous in the solar atmosphere, in particular in the intergranular space and in the chromosphere. One possible solution to overcome this problem is to adopt a more sophisticated identification method which is not affected by shear flows. In what follows, we review in detail the criterion that we consider most appropriate for this study: the swirling strength.


\subsection{Swirling strength}

The swirling strength criterion $\lambda$
\citep{1999JFM...387..353Z} 
is defined through the eigenanalysis of the velocity gradient tensor $\mathcal{U}$, 
\begin{equation}
  \mathcal{U}_{ij} \coloneqq \partial_j v_i    \,\,\leftrightarrow\,\, \mathcal{U}\equiv
    \begin{bmatrix}
      \partial_x v_x & \partial_y v_x & \partial_z v_x \\
      \partial_x v_y & \partial_y v_y & \partial_z v_y \\
      \partial_x v_z & \partial_y v_z & \partial_z v_z
  \end{bmatrix}\,. \label{eq:U_matrix}
\end{equation} 
This matrix is invariant under Galilean transformations and in one-to-one correspondence with vorticity via its anti-symmetrized version: $\vec{\omega} \leftrightarrow \mathcal{U} - \mathcal{U}^{\rm{T}}$. In general, when $\mathcal{U}$ has three discernible eigenvalues, they are either all real or one is real and there is a pair of complex conjugates \citep{doi:10.2514/6.1999-3288}. 

Whenever a vortex is present in the flow, the velocity gradient tensor $\mathcal{U}$ at that location is diagonalizable and can be decomposed in the following form,
\begin{eqnarray}
  \mathcal{U} =  
    \underbrace{\vphantom{\begin{bmatrix}
		                  \lambda_{\rm r} & 0 &  0\\
		                  0 & \lambda_{\rm +}& 0 \\
		                  0 & 0 & \lambda_{\rm -}  
		                  \end{bmatrix}} 
    \left[ \vec{u}_{\rm r}, \vec{u}_{\rm +}, \vec{u}_{\rm -}\right]
    }_{\textstyle\mathcal{P}}
	\underbrace{%
	\begin{bmatrix}
      \lambda_{\rm r} & 0 &  0\\
      0 & \lambda_{\rm +}& 0 \\
      0 & 0 & \lambda_{\rm -}  
    \end{bmatrix}}_{\textstyle\Lambda}
    \underbrace{\vphantom{\begin{bmatrix}
		                  \lambda_{\rm r} & 0 &  0\\
		                  0 & \lambda_{\rm +}& 0 \\
		                  0 & 0 & \lambda_{\rm -}  
		                  \end{bmatrix}} 
	\left[ \vec{u}_{\rm r}, \vec{u}_{\rm +}, \vec{u}_{\rm -}\right]^{-1}
	}_{\textstyle\mathcal{P}^{-1}}\,,  \label{eq:U_decomposition}
\end{eqnarray}
where $\vec{u}_{\rm r}, \vec{u}_{\rm +}$, and $\vec{u}_-$ are the eigenvectors of $\mathcal{U}$ associated with the eigenvalues $\lambda_{\rm r}, \lambda_{\rm +}$, and $\lambda_{\rm -}$. They form respectively the matrix of eigenvectors $\mathcal{P}$ (and its inverse $\mathcal{P}^{-1}$) and the matrix of the eigenvalues $\Lambda$. To provide more details, $\lambda_{\rm r}$ is a real eigenvalue, while $\lambda_{\rm +}$ and $\lambda_{\rm -}$ are complex conjugates, that is, $\lambda_{\pm} = \lambda_{\rm cr} \pm  \mathrm{i}\lambda_{\rm ci}$. In particular, their imaginary part, $\lambda_{\rm ci}$, characterizes the strength of a rigid body swirling motion through $T \coloneqq 2\pi/\lambda_{\rm ci}$.  It is therefore natural to define the swirling strength as $\lambda_{\rm ci}$, as done, for example, in 
\citet{2011A&A...533A.126M}. 
Nevertheless, in the present paper, we use a different convention; in fact, since our goal is to derive a dynamical equation for the swirling strength and then compare it to the vorticity equation, Eq.\,(\ref{eq:vorticity}), it is necessary that both criteria return the same value for a given vortex. Therefore, we decide to define the swirling strength as $\lambda = 2\lambda_{\rm ci}$. In this way, whenever the flow follows a rigid body rotation of period $T$, both criteria return the same value, $\lambda = \omega = 4\pi/T$. A more detailed description of the physical interpretation of $\lambda_{\rm r}$, $\lambda_{\rm cr}$, and $\lambda_{\rm ci}$ can be found in Appendix\,\ref{app:physical_interp}.

Another important piece of information we can obtain from Eq.\,(\ref{eq:U_decomposition}) is the vortex axis and its orientation. In fact, the eigenvector $\vec{u}_{r}$, associated with the real eigenvalue $\lambda_{\rm r}$, identifies the direction of the vortex axis. However, this is not sufficient to determine the orientation of the swirl, that is, whether the flow is turning in a clockwise or counter-clockwise fashion around this axis. This arbitrariness occurs because eigenvectors are defined up to a multiplicative constant only; therefore their orientation in space can be reverted. Nevertheless, there exists a simple method to ensure the physicality of its orientation, which consists in checking the handedness of the $\mathbb{R}^3$ basis which is formed by the eigenvectors of $\mathcal{U}$. In fact, one can prove (see Appendix\,\ref{app:physical_interp}) that if the basis is left-handed, then the rotation is counter-clockwise in the direction given by $\vec{u}_{\rm r}$, while if it is right-handed, it turns in a clockwise fashion. In practice, we force the eigenvectors to form a left-handed basis by multiplying $\vec{u}_{\rm r}$ by $-1$ where necessary. In this way, we ensure that all the rotations are counter-clockwise in the direction defined by $\vec{u}_{\rm r}$, as prescribed by the right-hand rule.

Once the orientation of $\vec{u}_{\rm r}$ is fixed, we can define the ``swirling vector'' as $\vec{\lambda} \coloneqq \lambda \vec{u_r}$ and therefore be able to extract the same amount of information from a flow as we can do with vorticity. However, when it comes to evolution, vorticity still represents the only viable way since its dynamical equation is known. In fact, there is no reference in the literature to a dynamical equation describing the evolution of the swirling strength and, therefore, a research on the generation of vortices using this criterion was not possible. In what follows, we derive the swirling strength evolution equation, enabling therefore the study of vortex evolution with this more sophisticated criterion.


\section{The swirling equation}\label{sec:swirlingequation}

In order to derive an evolution equation for the swirling strength $\lambda$ or, more generally, for the swirling vector $\vec{\lambda}$, we seek a tensor equation for $\mathcal{U}$. Then, using the diagonalization properties given in Eq.\,(\ref{eq:U_decomposition}), we isolate the equation relative to $\lambda_{\rm ci}$ obtaining therefore a ``swirling equation''. 

The starting point is the momentum equation, Eq.\,(\ref{eq:MHD_momentum_eq}), to which we apply the gradient operator $\nabla$ via the tensor product,
\begin{equation}
  \nabla\left(\frac{d\vec{v}}{dt}\right) = \nabla\left( -\frac{1}{\rho}\nabla(p_{\rm g} + p_{\rm m}) + \frac{1}{\rho}(\vec{B}\cdot\nabla)\vec{B} - \nabla\Phi  \right)\,, \label{eq:tensor_equation}
\end{equation}
where we define the tensor product between two vectors $\vec{a}$ and $\vec{b}$, $\mathcal{C} = \vec{a}\vec{b}$, as the matrix $\mathcal{C}$ whose components are given by $\mathcal{C}_{ij} = a_i b_j$.  
The resulting Eq.\,(\ref{eq:tensor_equation}) is therefore a tensor equation combining the different components of the gradient and the momentum equation. 

As a first step, we develop the left-hand side of Eq.\,(\ref{eq:tensor_equation}) by looking at a generic tensor component $( i,j )$,
\begin{align}
  \partial_i \left(\frac{{\rm d}v_j}{{\rm d}t}\right) &= \partial_i\left(\partial_t v_j + \sum_k (v_k \partial_k)\,v_j \right) \nonumber \\
  &= \partial_t \left(\partial_i v_j\right) + \sum_k (v_k \partial_k)\,\partial_i v_j + \sum_k (\partial_i v_k)\,(\partial_k v_j) \nonumber \\
  &= \partial_t \mathcal{U}_{ij}^{\rm{T}} + \sum_k (v_k\partial_k)\, \mathcal{U}_{ij}^{\rm{T}} + \sum_k \mathcal{U}_{kj}^{\rm{T}} \mathcal{U}_{ik}^{\rm{T}} \nonumber\\
  &= \frac{\rm d}{{\rm d}t}\mathcal{U}_{ij}^{\rm{T}} + (\mathcal{U}^2)_{ij}^{\rm{T}}\,, \label{eq:lhs_tensorequation} 
\end{align}
where we use the explicit form of the material derivative, Eq.\,(\ref{eq:mat_deriv}), and the definition of the velocity gradient tensor $\mathcal{U}$ given in Eq.\,(\ref{eq:U_matrix}). In this way, one can replace the velocity field in terms of $\mathcal{U}$. Therefore, Eq.\,(\ref{eq:tensor_equation}) can be rewritten in the following form,
\begin{equation}
  \frac{\rm d}{{\rm d}t}\mathcal{U}^{\rm{T}} = -(\mathcal{U}^2)^{\rm{T}} + \mathcal{M}^{\rm{T}}\,,\nonumber
\end{equation}
and by simply transposing, 
\begin{equation}
  \frac{\rm d}{{\rm d}t}\mathcal{U} = -\mathcal{U}^2 + \mathcal{M}\,,\label{eq:U_evolution}
\end{equation}
where $\mathcal{M}$ is the transpose of the tensor obtained in the right-hand side of Eq.\,(\ref{eq:tensor_equation}) by applying the gradient to the source terms. To be more specific, a generic component of $\mathcal{M}$ is given by
\begin{equation}
 \mathcal{M}_{ij} = \partial_j\left( -\frac{1}{\rho}\partial_i\left(p_{\rm g} + p_{\rm m}\right) + \frac{1}{\rho}\left(\sum_k B_k \partial_k \right)B_i - \partial_i\Phi  \right)\,. \label{eq:M_tensor}
\end{equation} 

Successively, we use the diagonalization properties of $\mathcal{U}$ in order to transform Eq.\,(\ref{eq:U_evolution}) into an evolution equation for $\Lambda$. In particular, by inverting  Eq.\,(\ref{eq:U_decomposition}) one finds the usual transformation law for matrices,
\begin{equation}
  \Lambda = \mathcal{P}^{-1}\mathcal{U}\mathcal{P}\,, \label{eq:tensor_transformation}
\end{equation}
which leads to the following relations,
\begin{equation}
  \mathcal{P} \Lambda = \mathcal{U} \mathcal{P}\,, \quad 
  \Lambda \mathcal{P}^{-1} = \mathcal{P}^{-1} \mathcal{U}\,. \label{eq:tensor_relations}
\end{equation}
Moreover, given that $\mathcal{P}\mathcal{P}^{-1} = \mathcal{I}$, where $\mathcal{I}$ is the identity matrix, and that ${\rm d}\mathcal{I}/{\rm d}t = 0$, one can easily prove the following relation,
\begin{equation}
	\frac{\rm d}{{\rm d}t}(\mathcal{P}^{-1})\mathcal{P} = - \mathcal{P}^{-1}\frac{\rm d}{{\rm d}t}(\mathcal{P})\,. \label{eq:P_relation}
\end{equation}

Then, multiplying Eq.\,(\ref{eq:U_evolution}) by $\mathcal{P}^{-1}$ on the left and by $\mathcal{P}$ on the right we obtain,
\begin{equation}
	\mathcal{P}^{-1}(\frac{\rm d}{{\rm d}t}\mathcal{U})\mathcal{P} = -\mathcal{P}^{-1}\mathcal{U}^2\mathcal{P} + \mathcal{P}^{-1}\mathcal{M}\mathcal{P}\,. \label{eq:Lambda_eq_begin}
\end{equation}
Using Eqs.\,(\ref{eq:tensor_transformation}, \ref{eq:tensor_relations}, \ref{eq:P_relation}), for the left-hand side of Eq.\,(\ref{eq:Lambda_eq_begin}), we find, 
\begin{align}
	\mathcal{P}^{-1}(\frac{\rm d}{{\rm d}t}\mathcal{U})\mathcal{P} \,=&\, \frac{\rm d}{{\rm d}t}\left( \mathcal{P}^{-1}\mathcal{U}\mathcal{P} \right) - \frac{\rm d}{{\rm d}t}\left( \mathcal{P}^{-1} \right)\mathcal{U}\mathcal{P} - \mathcal{P}^{-1}\mathcal{U}\frac{\rm d}{{\rm d}t}\mathcal{P}\,, \nonumber\\
	=&\,\frac{\rm d}{{\rm d}t}\Lambda - \frac{\rm d}{{\rm d}t}\left(\mathcal{P}^{-1}\right)\mathcal{P}\Lambda - \Lambda\mathcal{P}^{-1}\frac{\rm d}{{\rm d}t}\mathcal{P}\,, \nonumber\\
	=&\, \frac{\rm d}{{\rm d}t}\Lambda - \Lambda\mathcal{P}^{-1}\frac{\rm d}{{\rm d}t}\mathcal{P} + \mathcal{P}^{-1}\frac{\rm d}{{\rm d}t}\left(\mathcal{P}\right) \Lambda\,, \nonumber\\
	=&\, \frac{\rm d}{{\rm d}t}\Lambda - \left[ \Lambda, \mathcal{P}^{-1}\frac{\rm d}{{\rm d}t}\left(\mathcal{P}\right) \right]\,, \nonumber
\end{align}
where the square brackets indicate the commutator of two matrices,  $\left[\mathcal{A},\mathcal{B} \right] \coloneqq \mathcal{A}\mathcal{B}-\mathcal{B}\mathcal{A}$. For the right-hand side instead, we get,
\begin{align*}
	-\mathcal{P}^{-1}\mathcal{U}^2\mathcal{P} + \mathcal{P}^{-1}\mathcal{M}\mathcal{P} \,=&\, -\mathcal{P}^{-1}\mathcal{U}\mathcal{I}\mathcal{U}\mathcal{P} + \mathcal{P}^{-1}\mathcal{M}\mathcal{P}\,,\\
	=&\,-\mathcal{P}^{-1}\mathcal{U}\mathcal{P}\,\mathcal{P}^{-1}\mathcal{U}\mathcal{P} + \mathcal{P}^{-1}\mathcal{M}\mathcal{P}\,,\\
	=&\, -\Lambda^2 + \mathcal{P}^{-1}\mathcal{M}\mathcal{P}\,. \nonumber
\end{align*}
Combining back together the two sides of Eq.\,(\ref{eq:Lambda_eq_begin}), we find an equation which couples the evolution of the eigenvalue matrix $\Lambda$ with the eigenvector matrix $\mathcal{P}$,
\begin{equation}
	\frac{\rm d}{{\rm d}t}\Lambda = \left[ \Lambda, \mathcal{P}^{-1}\frac{\rm d}{{\rm d}t}\left(\mathcal{P}\right) \right] - \Lambda^2 + \mathcal{P}^{-1}\mathcal{M}\mathcal{P}\,. \label{eq:Lambda+P_eq}
\end{equation}
However, this coupling is just apparent since the commutator between a diagonal matrix and any other matrix results in an hollow matrix. The interested reader can refer to Appendix\,\ref{app:determinant_proof} for a proof.
Consequently, the commutator in Eq.\,(\ref{eq:Lambda+P_eq}) between $\Lambda$ and $\mathcal{P}^{-1}{\rm d}\mathcal{P}/{\rm d}t$ is hollow, thus it contributes to the tensor equations only in the off-diagonal terms. 
Since the total derivative of $\Lambda$ is restricted to the diagonal terms of Eq.\,(\ref{eq:Lambda+P_eq}), the evolution of $\Lambda$ and $\mathcal{P}$ is decoupled and we can easily distinguish two sets of equations relative to each quantity. 
For the eigenvalues matrix $\Lambda$, we consider alone the equations relative to the diagonal terms,
\begin{equation}
   \frac{\rm d}{{\rm d}t}\Lambda_{ii} = -\left(\Lambda^2\right)_{ii} + \left(\mathcal{P}^{-1}\mathcal{M}\mathcal{P}\right)_{ii}\,, \label{eq:Lambda_eq} 
\end{equation}
where the index $i$ refers to the $i$-th eigenvalue of $\Lambda$,
while for the eigenvector matrix $\mathcal{P}$, we take into account the off-diagonal terms,
\begin{equation}
    \left[ \Lambda, \mathcal{P}^{-1}\frac{\rm d}{{\rm d}t}\left(\mathcal{P}\right) \right]_{ij} + \left(\mathcal{P}^{-1}\mathcal{M}\mathcal{P}\right)_{ij} = 0 \,,\quad i \neq j\,. \label{eq:P_eq}
\end{equation}
Since we are looking for a dynamical equation for the swirling vector, $\vec{\lambda} = \lambda\vec{u}_{\rm r}$, we are left with the extraction of the evolution equations for the swirling strength $\lambda$ and the vortex axis $\vec{u}_{\rm r}$ from Eqs.\,(\ref{eq:Lambda_eq}, \ref{eq:P_eq}).

The evolution of the three eigenvalues present in $\Lambda$ is dictated by the relative equation given in Eq.\,(\ref{eq:Lambda_eq}). Because of the definition of swirling strength, we can restrict ourselves to the one relative to the eigenvalue $\lambda_+ = \lambda_{\rm cr} + \rm{i}\lambda_{\rm ci}$,
\begin{align}
    \frac{\rm d}{{\rm d}t}\lambda_+ &= - \lambda_+^2 + \left(\mathcal{P}^{-1}\mathcal{M}\mathcal{P}\right)_{22}\,, \nonumber \\
    \frac{\rm d}{{\rm d}t}\left(\lambda_{\rm cr} + \rm{i} \lambda_{\rm ci}\right) &= -\left( \lambda_{\rm cr} + \rm{i} \lambda_{\rm ci}\right)^2 + \left(\mathcal{P}^{-1}\mathcal{M}\mathcal{P}\right)_{22}\,\nonumber. 
\end{align}
By taking only the imaginary part, we recover an equation for $\lambda_{\rm ci}$. Then, given our convention, we formulate the swirling equation as,
\begin{align}
    \frac{\rm d}{{\rm d}t}\lambda =& -2\lambda\lambda_{\rm cr} + 2 \rm{Im}\left( \mathcal{P}^{-1}\mathcal{M}\mathcal{P} \right)_{22}\,, & &\nonumber\\
    =& \vphantom{\left\{ \mathcal{P}^{-1} \left[ \nabla \bigg(  \frac{1}{\rho}\nabla p_{\rm g} \bigg) \right] \mathcal{P} \right\}}
    - 2\lambda\lambda_{\rm cr} & T^1_{\rm \lambda} & \nonumber \\
    & - 2\rm{Im}\left\{ \mathcal{P}^{-1} \left[ \nabla \bigg(  \frac{1}{\rho}\nabla \textit{p}_{\rm g} \bigg) \right] \mathcal{P} \right\}_{22} & T^2_{\rm \lambda} & \nonumber \\
    & - 2\rm{Im}\left\{ \mathcal{P}^{-1} \left[ \nabla \bigg(  \frac{1}{\rho} \nabla  \textit{p}_{\rm m} \bigg) -  \bigg(\nabla\frac{1}{\rho} \bigg)  (\vec{B}\cdot\nabla) \vec{B}  \right] \mathcal{P} \right\}_{22} & T^3_{\rm \lambda} & \nonumber \\
    & + 2\rm{Im}\left\{ \mathcal{P}^{-1} \left[ \frac{1}{\rho} \nabla \bigg( \Big( \vec{B} \cdot \nabla \Big) \vec{B} \bigg) \right] \mathcal{P} \right\}_{22} & T^4_{\rm \lambda} & \nonumber \\
    & \vphantom{\left\{ \mathcal{P}^{-1} \left[ \nabla \bigg(  \frac{1}{\rho}\nabla p_{\rm g} \bigg) \right] \mathcal{P} \right\}} 
    - 2\rm{Im}\left\{ \mathcal{P}^{-1} \bigg[ \nabla \Big( \nabla \Phi \Big) \bigg] \mathcal{P} \right\}_{22}\,. & T^5_{\rm \lambda} 
    &    \label{eq:swirling_eq}
\end{align}

In order to give a physical interpretation to Eq.\,(\ref{eq:swirling_eq}), we adopt a term labeling similar to the one in the vorticity equation, Eq.\,(\ref{eq:vorticity}). In particular, we try to match terms that share the same physical mechanism for vortex generation. Therefore, $T^{1}_{\rm \lambda}$ can be considered as a stretching term, since it involves $\lambda_{\rm cr}$, which indicates compression or dilatation of the flow in the rotation plane, as we show in Appendix \ref{app:physical_interp}. The second term, $T^2_{\rm \lambda}$, is related to the hydrodynamic baroclinicity, while $T^3_{\rm \lambda}$ is its magnetic correspondent. The generation of swirling strength by magnetic tension is instead represented by the term $T^4_{\rm \lambda}$. Finally, we note that the last term, $T^5_{\rm \lambda}$, which is associated to the potential of conservative forces, has no analog in the vorticity equation. When $\Phi$ represents the gravitational potential and we can assume a slowly varying unidirectional gravitation field, as it is the case in most simulations of the solar atmosphere, $T^5_{\rm \lambda}$ can be safely neglected. However, this term could be relevant whenever the configuration of the gravitational potential is complex and inhomogeneous.

The equation derived for the swirling strength, Eq.\,(\ref{eq:swirling_eq}), is similar in form to the evolution equation of vorticity, Eq.\,(\ref{eq:vorticity}). The main difference consists in the multiplication of each source term present in $\mathcal{M}$ by $\mathcal{P}^{-1}$ and $\mathcal{P}$. One can physically interpret this difference recalling that the eigenvectors that compose $\mathcal{P}$ form an eigenbasis of the velocity gradient tensor $\mathcal{U}$. In such a basis, $\mathcal{U}$ explicitly describes the characteristics of the flow through its eigenvalues. 
Therefore, $\mathcal{P}^{-1} \mathcal{M} \mathcal{P}$ represents a change of basis of the source terms, and the basis defined by $\mathcal{P}$ describes the proper coordinate system of the vortex, where the shears and the other components of the flow, which could mislead the vortex interpretation, have been implicitly taken into account in the basis vectors. 
Consequently, the terms appearing in Eq.\,(\ref{eq:swirling_eq}) are related to the generation and the evolution of true vortices only. 

Finally, we note that we could have chosen to pick the equation relative to the other complex conjugate eigenvalue, $\lambda_{\rm -} = \lambda_{\rm cr} - \rm{i}\lambda_{\rm ci}$, for the formulation of the swirling equation. This leads to an equation which, at first sight, appears different, but in reality is equivalent to Eq.\,(\ref{eq:swirling_eq}). A proof is given in Appendix\,\ref{app:alternative_deriv}.
Moreover, one can also derive evolution equations for the $\lambda_{\rm cr}$ and $\lambda_{\rm r}$ parameters in a similar fashion.

Given the swirling equation, we are left with deriving an equation for the evolution of the eigenvector $\vec{u}_{\rm r}$. 
To do so, we turn to Eq.\,(\ref{eq:P_eq}) using the property of commutators derived in Eq.\,(\ref{eq:commutator_proof}),
\begin{align}
\left[  \mathcal{P}^{-1}\frac{\rm d}{{\rm d}t}\left(\mathcal{P}\right), \Lambda \right]_{ij} &=  \left( \mathcal{P}^{-1}\mathcal{M}\mathcal{P} \right)_{ij} \,, \nonumber \\
\left( \lambda^{j} - \lambda^{i} \right) \left( \mathcal{P}^{-1}\frac{\rm d}{{\rm d}t}\mathcal{P} \right)_{ij} &=  \left( \mathcal{P}^{-1}\mathcal{M}\mathcal{P} \right)_{ij} \,,\nonumber \\
\left( \mathcal{P}^{-1} \frac{\rm d}{{\rm d}t} \mathcal{P} \right)_{ij} &= \frac{1}{ \left( \lambda^{j} - \lambda^{i} \right) } \left(\mathcal{P}^{-1} \mathcal{M} \mathcal{P} \right)_{ij} \,, \nonumber \\
\sum_k \mathcal{P}^{-1}_{ik}\frac{\rm d}{{\rm d}t} P_{kj} &= \frac{1}{ \left( \lambda^{j} - \lambda^{i} \right) } \left( \mathcal{P}^{-1} \mathcal{M} \mathcal{P} \right)_{ij} \,, \,i \neq j \,,\label{eq:eigenvector_eq} 
\end{align}
where in the second equation we use the fact that $\left[\mathcal{A}\,,\mathcal{B}\right] = -\left[\mathcal{B}\,,\mathcal{A}\right] $ and in the last equation we explicitly multiply $\mathcal{P}^{-1}$ with $d\mathcal{P}/dt$ and we recall that this equation is valid only for $i \neq j$. 

Equation\,(\ref{eq:eigenvector_eq}) consists of six evolution equations, for all combinations of $i$ and $j$ such that $i\neq j$, where $j$ indexes the evolving eigenvector. Since we are interested in the evolution of $\vec{u}_{\rm r}$, we concentrate on the $j=1$ components of Eq.\,(\ref{eq:eigenvector_eq}),
\begin{equation}
  \sum_k \mathcal{P}^{-1}_{ik}\frac{\rm d}{{\rm d}t} \left( \vec{u}_{\rm r} \right)_k = \frac{1}{\left(  \lambda_{\rm r} - \lambda^{i}\right)} \left( \mathcal{P}^{-1} \mathcal{M} \mathcal{P} \right)_{i1} \,,\quad i = 2,3\,,\label{eq:u_r_equation}
\end{equation}
where we use the fact that $\lambda^1 = \lambda_{\rm r}$ in our diagonalization convention of the velocity gradient tensor $\mathcal{U}$. Equation (\ref{eq:u_r_equation}) consists of two equations, coupling the evolution of the three components of the eigenvector $\vec{u}_{\rm r}$. We miss one equation in order to solve the system. One simple extra constraint comes from the normalization of $\vec{u}_{\rm r}$, which we fix to one, 
$(\vec{u}_{\rm r})_x^2 + (\vec{u}_{\rm r})_y^2 + (\vec{u}_{\rm r})_z^2 =1$, which can be expressed as
\begin{equation}
  (\vec{u}_{\rm r})_x \frac{\rm d}{{\rm d}t} (\vec{u}_{\rm r})_x +  (\vec{u}_{\rm r})_y \frac{\rm d}{{\rm d}t} (\vec{u}_{\rm r})_y +  (\vec{u}_{\rm r})_z \frac{\rm d}{{\rm d}t} (\vec{u}_{\rm r})_z = 0\,. \nonumber
\end{equation}
Thus, we have now a set of three equations that describe the dynamical evolution of the vortex axis.

Combining the results obtained, we can describe the evolution of the swirling vector $\vec{\lambda}$ by
\begin{equation}
    \frac{\rm d}{{\rm d}t}\vec{\lambda} = \vec{u}_{\rm r}\frac{\rm d}{{\rm d}t}\lambda + \lambda\frac{\rm d}{{\rm d}t}\vec{u}_{\rm r} \,, \label{eq:totalswirlingeq}
\end{equation}
where the first term represents the generation of swirling strength while the second characterizes the evolution of the vortex axis. We could therefore say that the term $\lambda {\rm d} \vec{u}_{\rm r}/{\rm d}t$ in Eq.\,(\ref{eq:totalswirlingeq})
is the analog of the tilting term, $T^0_{\rm \lambda}$, present in Eq.\,(\ref{eq:vorticity}). However, in the present case, this term is not responsible for any vortex generation, but instead it describes the transition under constant swirling strength between components of $\vec{\lambda}$ when a vortex tilts its orientation.


\section{Results and discussion}\label{sec:results}

In this section, we compare the generation of vorticity and swirling strength, in particular with regard to their evolution equations, Eqs.\,(\ref{eq:vorticity}) and (\ref{eq:swirling_eq}). Firstly, we  study two simple scenarios to better understand the limits of vorticity and to highlight the advantages of having an equation for the swirling strength available. Then, we apply the dynamical equations to high-resolution radiative magneto-hydrodynamic simulation data produced with the CO$^5$BOLD code \citep{2012JCoPh.231..919F} 
to study the mechanisms of vortex generation. The simulation was carried out by starting from a previous, relaxed hydrodynamical model to which a uniform vertical magnetic field with a strength of $50\,\mathrm{G}$ was superimposed. This magnetic model was then further advanced with the MHD module of CO$^5$BOLD, which uses an HLL approximate Riemann solver \citep{doi:10.1137/1025002} 
until relaxation of the system is achieved. The lateral boundary conditions are periodic, while at the top and at the bottom they are open under the condition that the net mass flux at the bottom boundary vanishes. The magnetic field is constrained to be vertical at both top and bottom boundaries but it can freely move in lateral directions. 

The Cartesian computational domain encompasses a volume of a $9.6\,\times\,9.6\,\times\,2.8\,\mathrm{Mm}^3$ of the solar atmosphere, ranging from the surface layers of the convection zone to the chromosphere. The optical surface $\tau_{500} = 1$ is approximately located in the middle of the height range. The size of the computational cells is $10\,\times\,10\,\times\, 10\,\mathrm{km}^3$, uniform in the entire box.
The gravitational field is vertical and uniform with a constant value of $\log(g) = 4.44$. More details on this simulation can be found in 
\citet{doi:10.13097/archive-ouverte/unige:115257}. 
Here we use a sequence over a time period of about
$10\,\mathrm{min}$ real time of solar evolution for which we have time instants
stored every $10\,\mathrm{s}$.

For the present study we consider vertical vortices only, that is, those for which the flow rotates around the $z$-coordinate axis,
$\hat{z}$. 
We are particularly interested in this kind of vortices since they potentially transport energy and mass across different layers of the solar atmosphere. Nevertheless, we would like to point out that the method is suitable to all kind of vortices.


\subsection{Limits of vorticity}\label{subsec:analytical_approach}

\begin{figure*}
    \centering
    \includegraphics[height=7cm]{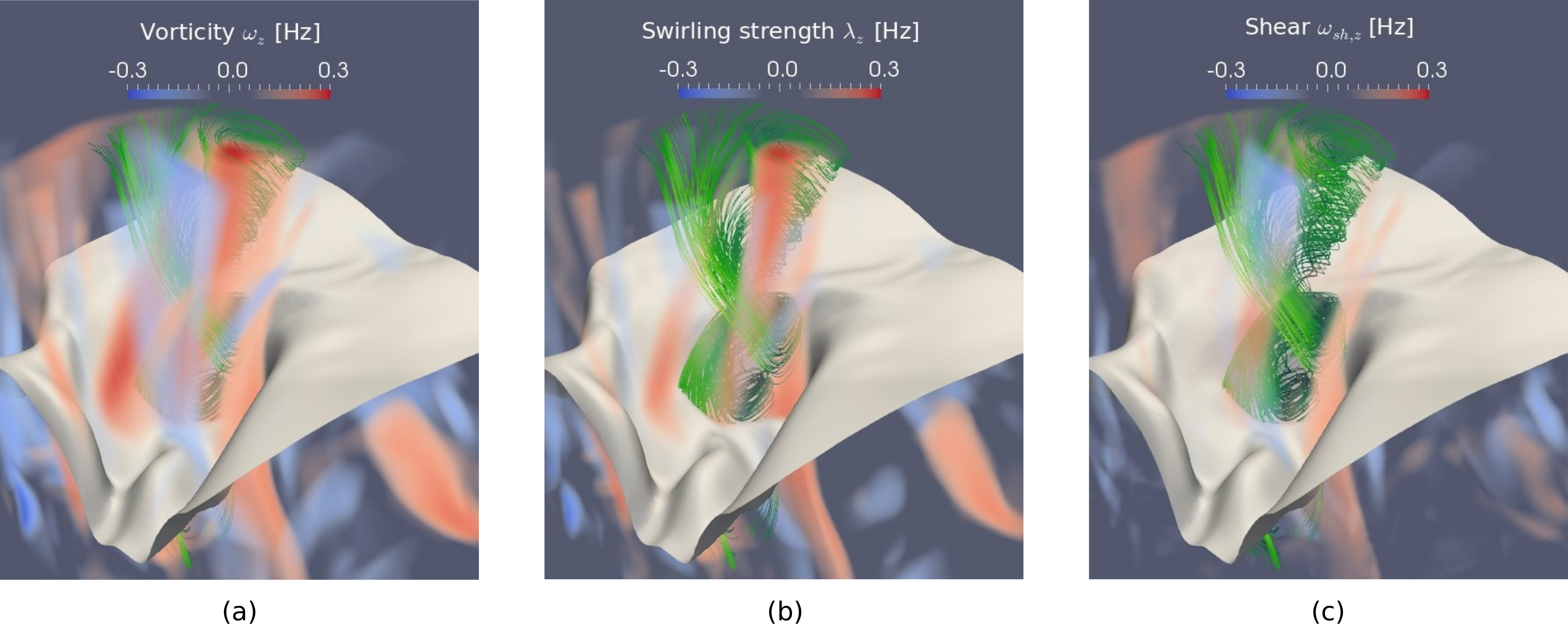}
	\caption{Vertical component of (a) vorticity, (b) swirling strength, and (c) shear strength, computed in a small portion of the full simulation domain, measuring $0.6\,\times\,0.6\,\times\,1.8\, \mathrm{Mm}^3$ and encompassing an intergranular region with strong vertical magnetic field ($B_z \gtrsim 500$ G). The gray sheet displays the surface of optical depth $\tau_{500} = 1$, while the plasma flow is rendered with green streamlines.
	These images were produced with the software ParaView \citep{Ahrens+al2005}. 
	}
	\label{fig:shearflows}
\end{figure*}

We start with a simple rotational flow with velocity $\vec{v}$ and corresponding vorticity $\vec{\omega}$ given by,
\begin{equation}
    \vec{v} = \begin{pmatrix}
            -y\frac{\Omega}{2} \\[2pt]
             \phantom{-}x\frac{\Omega}{2} \\[2pt]
             0
        \end{pmatrix}
    \,,\qquad
    \vec{\omega} = \nabla \times \vec{v} = \begin{pmatrix}
            0 \\
            0 \\
            \Omega
        \end{pmatrix}
    \,. \nonumber 
\end{equation}
In this example, the flow is performing a rigid, counter-clockwise rotation of period $T = 4\pi/\Omega$. Without loss of generality, we have chosen the axis of rotation to be along the $\hat{z}$-direction.
Then, we can also compute the velocity gradient tensor $\mathcal{U}$ and, by its diagonalization, find the swirling strength,
\begin{eqnarray}
    \mathcal{U} &=& 
        \begin{bmatrix} 
          0                   & -\frac{\Omega}{2} & 0 \\[2pt]
          \frac{\Omega}{2}    &   0               & 0 \\[2pt]
          0                   &   0               & 0 
        \end{bmatrix}\,, \nonumber\\
        &=& 
        \begin{bmatrix} 
          0  & -{\rm i}\frac{1}{\sqrt{2}} & {\rm i}\frac{1}{\sqrt{2}} \\[5pt]
          0  & -\frac{1}{\sqrt{2}}        & -\frac{1}{\sqrt{2}} \\[5pt]
          1  &       0                    & 0 
        \end{bmatrix} 
        \begin{bmatrix} 
          0 &     0                    &   0 \\[5pt]
          0 & {\rm i}\frac{\Omega}{2}  &   0 \\[5pt]
          0 &     0                    & -{\rm i}\frac{\Omega}{2}      
        \end{bmatrix}
        \begin{bmatrix} 
          0                          & 0                     & 1 \\[2pt]
          {\rm i}\frac{1}{\sqrt{2}}  & -\frac{1}{\sqrt{2}}   & 0 \\[5pt]
          -{\rm i}\frac{1}{\sqrt{2}} & -\frac{1}{\sqrt{2}}   & 0 
        \end{bmatrix}\,, \nonumber\\
        &=& \mathcal{P} \Lambda \mathcal{P}^{-1}\,. \label{eq:rotational_example}
\end{eqnarray}
Given our convention, we see that $\lambda \equiv 2\lambda_{\rm ci} = \Omega$. Moreover, by computing the real matrix $\Tilde{\mathcal{P}}$ (defined in Appendix \ref{app:physical_interp}) and its determinant, one can check 
that the $\mathbb{R}^3$ basis in which the swirl is described is left-handed. Thus, we conclude that the vortex axis is given by $\vec{u}_{\rm r} = (0,0,1)$ and we find that, for this simple example, the swirling and the vorticity vectors are identical, $\vec{\lambda} = \vec{\omega} = (0, 0, \Omega)$. This should not surprise us: indeed, for purely rotational flows, vorticity and swirling strength should be equivalent given the absence of shear flows. 

At this point, we add a shear flow along the $\hat{z}$-axis proportional to the $\hat{y}$ coordinate in such a way that the velocity and vorticity vectors are,
\begin{equation}
    \vec{v} = \begin{pmatrix}
            -y\frac{\Omega}{2} \\[2pt]
            x\frac{\Omega}{2} \\[2pt]
            y\xi
        \end{pmatrix}
    \,,\qquad
    \vec{\omega} = \nabla \times \vec{v} = \begin{pmatrix}
            \xi \\
            0 \\
            \Omega
        \end{pmatrix}
    \,, \nonumber 
\end{equation}
where $\xi$ is the strength of the shear. We notice that the extra component in the velocity vector, $\vec{v}_z = y\xi$, which is not related to any kind of rotation in the flow, generates vorticity along the $\hat{x}$-axis. 
Computing once again the velocity gradient tensor and its diagonal decomposition we find,
\begin{eqnarray}
  \mathcal{U}  &=&
    \begin{bmatrix} 
        0                   & -\frac{\Omega}{2} & 0 \\[2pt]
        \frac{\Omega}{2}    &   0               & 0 \\[2pt]
        0                   &   \xi             & 0 
    \end{bmatrix}\,, \nonumber\\
    &=&  
    \begin{bmatrix} 
        0   & -\frac{\Omega}{2\xi c}        & -\frac{\Omega}{2\xi c} \\[4pt]
        0   & \rm{i}\frac{\Omega}{2\xi c}  & -\rm{i}\frac{\Omega}{2\xi c} \\[4pt]
        1   &       \frac{1}{c}          & \frac{1}{c} 
    \end{bmatrix} 
    \begin{bmatrix} 
        0   &     0                    &   0 \\[5pt]
        0   & {\rm i}\frac{\Omega}{2}  &   0 \\[5pt]
        0   &     0                    & -{\rm i}\frac{\Omega}{2}      
    \end{bmatrix}
    \begin{bmatrix} 
        \,\,\frac{2\xi}{\Omega}  & \,\,0                            & 1 \\[3pt]
        -\frac{\xi c}{\Omega}  & -{\rm i} \frac{\xi c}{\Omega}  & 0 \\[3pt]
        -\frac{\xi c}{\Omega}  & \,\,{\rm i} \frac{\xi c}{\Omega} & 0 
    \end{bmatrix} \,, \label{eq:rotational+shear_example}
\end{eqnarray}
where $c = \sqrt{1 + \Omega^2/(2\xi^2)}$. Computing the matrix of real eigenvectors, $\Tilde{\mathcal{P}}$, and checking the handedness of the corresponding basis according to Appendix \ref{app:physical_interp}, one can conclude that the swirling strength vector $\vec{\lambda} = (0,0,\Omega)$ is not affected by the shear flow and correctly describes the properties of the vortex. One can therefore define the shear strength, $\vec{\omega}_{sh} \coloneqq \vec{\omega} - \vec{\lambda}$, as the contribution to vorticity due to shear flows. In practice, the shear strength is a proxy for estimating how vorticity is biased by the shears present in the flow. Therefore, it can indicate the regions of the plasma where vorticity cannot be trusted for the identification of vortex flows.

These two simple examples are useful to get familiar with the criteria used to study vortices in MHD. However, when it comes to study realistic scenarios, the plasma dynamics are much more complex. 
Panels (a) and (b) of Fig.\,\ref{fig:shearflows} illustrate the vertical components of vorticity $\vec{\omega}_z$ and swirling strength $\lambda_z$ for one particular vortex in an intergranular region with a strong vertical magnetic field. We also show in gray shades the optical surface of $\tau_{500}=1$.
The presence of the vortex is highlighted by the streamlines (in green), which indicate rotation around the strong concentration of positive swirling strength and vorticity (red). However, we notice how the swirling strength is predominately positive and how, above the optical surface, it is mainly concentrated inside the vortex, while vorticity presents also negative filaments (blue) and looks slightly more dispersed. Considering vorticity alone, one could be lead to conclude that panel (a) of Fig.\,\ref{fig:shearflows} shows two vortices counter-rotating one with respect to the other. However, it is clear from panel (b) of Fig.\,\ref{fig:shearflows} and from the streamlines that the negative vortex hinted by the vorticity is, in reality, a thin sheath of shear flow. As  can be seen from panel (c) of Fig.\,\ref{fig:shearflows}, the shears are all-over the atmospheric portion of interest, thus the importance of the swirling strength approach.


\subsection{Comparing the generation of vorticity and 
swirling strength from their respective equations}
\label{subsec:vortvsswirl}

\begin{figure}
	\centering
	\resizebox{\hsize}{!}{\includegraphics{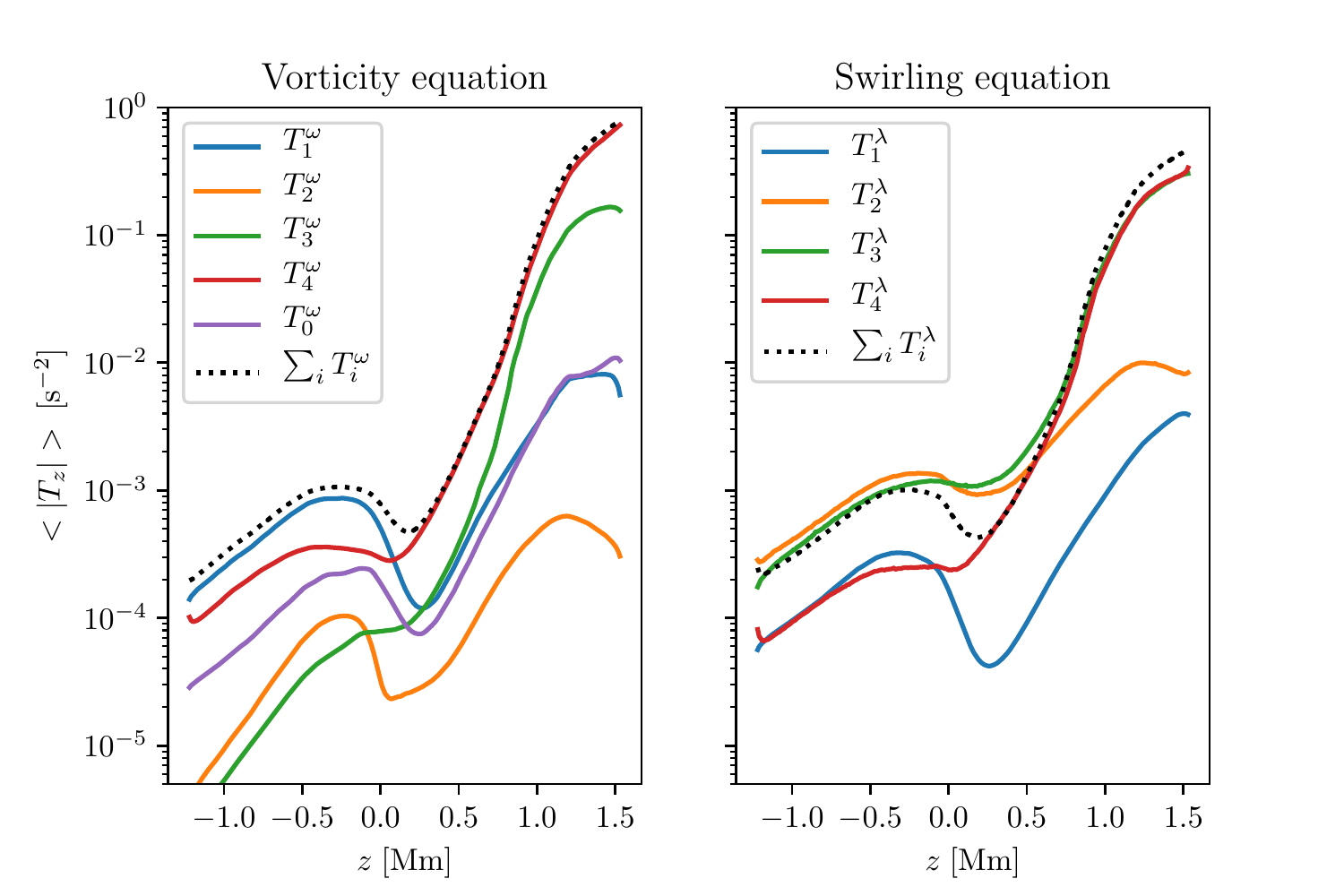}}
	\caption{Mean absolute values of the vertical components of the terms appearing on the right-hand side of the vorticity equation (left panel) and the swirling equation (right panel) as a function of $z$. The dotted curves represent the mean absolute value of the total production of vorticity and swirling strength in the simulation. The optical surface $\tau_{500} = 1$ corresponds to $z=0$.}
	\label{fig:terms_all}
\end{figure}

This section investigates the generation of vortical flows in the solar atmosphere using the evolution equations of vorticity, Eq.\,(\ref{eq:vorticity}), and of the swirling vector, Eq.\,(\ref{eq:totalswirlingeq}). In particular, we want to study which physical process is dominant at different heights in the atmosphere, from the upper convection zone through the photosphere up to the chromosphere, and compare the results between the two approaches. For this purpose, we study the first term of Eq.\,(\ref{eq:totalswirlingeq}) alone, since we are interested in the generation of the vertical component of vortices only and not in their tilting from one axis to another.
In particular, $\vec{u}_{\rm r}{\rm d}\lambda/{\rm d}t$ is related to the generation of swirling strength through Eq.\,(\ref{eq:swirling_eq}). Therefore, in what follows, we define the vertical component of the swirling equation terms, $T_{i,z}^{\lambda}$, as being the $\hat{z}$-component of the generating part of Eq.\,(\ref{eq:totalswirlingeq}),
\begin{equation}
    (\vec{u}_{\rm{r}})_z \frac{\rm{d}}{{\rm d}t}\lambda = (\vec{u}_{\rm{r}})_z\sum_i T_{i}^{\lambda} = \sum_i T_{i,z}^{\lambda}\,.\nonumber
\end{equation}

Figure\,\ref{fig:terms_all} shows the mean absolute values of the vertical components of the terms appearing in Eqs.\,(\ref{eq:vorticity}) and (\ref{eq:swirling_eq}) as a function of the height $z$. Moreover, we note that $T_5^{\rm \lambda}$ is neglected in this study because the present simulation has a constant gravitational acceleration and, consequently, $\nabla(\nabla\Phi)=0$.

Concerning vorticity, we notice that the tilting term $T_1^{\omega}$ is the largest contributor from the convection zone up to the low photosphere, where it gets overtaken by the magnetic tension term $T_4^{\omega}$. From there on further up, magnetic effects are mainly responsible for the generation of vorticity. In particular, we see that in the chromospheric layers, the sum of all components is almost identical to the magnetic tension term alone, implying that the other terms are minor in this height range. We also notice a drop in vorticity generation in the photosphere, which can be ascertained to the steep drop in density at this height with the consequence that up-flows expand. This results in a loss of rotational speed because of angular momentum conservation
\citep{1997A&A...328..229N}. 

A similar study has already been worked out by 
\citet{2011A&A...526A...5S}, 
who used the MURaM code 
\citep{Vogler2003, 2005A&A...429..335V} 
to simulate the magnetized photosphere. We find very similar results to the ones presented there, apart from what concerns the terms $T_1^{\omega}$ and $T_2^{\omega}$, which seem to have been interchanged by \citet{2011A&A...526A...5S}. 
This conclusion is also supported by the results of \citet{keerthana+al2020}, 
which also show the presence of a strong tilting term in the upper convection zone and a subdominant hydrodynamical baroclinic term in simulations performed with the MURaM, CO$^5$BOLD, BIFROST \citep{2011A&A...531A.154G}, 
and STAGGER 
\citep{1994ASIC..433..471N, nordlund+galsgaars1997} 
codes.

Yet, we argue that the results obtained with vorticity can be biased, since its equation also accounts for the generation of shears. Therefore, the right panel of Fig.\,\ref{fig:terms_all} presents the same analysis using the swirling equation. In this case, we see that the stretching term, $T_1^{\lambda}$, is unimportant over the entire height range. In the convection zone, the two main contributors to the generation of swirling strength are the hydrodynamical baroclinic term $T_2^{\lambda}$ and the magnetic baroclinic term $T_3^{\lambda}$, while in the chromosphere we confirm that the magnetic terms, $T_3^{\lambda}$ and $T_4^{\lambda}$, are by far the most important ones. However, for the generation of vorticity, the effects of magnetic tension  are stronger than those of magnetic pressure, while in the right panel of Fig.\,\ref{fig:terms_all} we see that these two magnetic terms have almost identical mean contributions. In the photosphere, we observe a decrease in the total production of swirling strength for the same reason that leads to a drop in vorticity generation. 

Interestingly, the mean total generation of swirling strength in the superficial layer of the convection zone and, in particular in the photosphere, is lower that the mean values of the plasma and magnetic baroclinic terms, implying that these two terms partially counter-balance their effects. This is intuitive if we think of a magnetic flux tube, where the external gas pressure must balance the sum of internal magnetic and gas pressures 
\citep{1978SoPh...60..213Z}. 
Thus, a positive gradient in magnetic pressure opposes a negative one in gas pressure, which results in opposite contributions to the baroclinic terms $T^{\lambda}_2$ and $T^{\lambda}_3$ in
Eq.\,(\ref{eq:swirling_eq}).
On the other hand, this interpretation implies that the most important contributor to vortex generation in the photosphere and in the low chromosphere is the magnetic tension term, $T_4^{\lambda}$.

In conclusion, Fig.\,\ref{fig:terms_all} tells us that there are substantial differences between the two approaches to identification of vortex generation, in particular regarding the convection zone. Obviously, these differences are caused by the generation of shears along with vortices, which is included in the right-hand side terms of the vorticity equation and which can lead into erroneous analysis. Therefore, we think that the best tool to study vortex generation in the solar atmosphere is the swirling equation presented in Sect.\,\ref{sec:swirlingequation}.  Furthermore, we can deduce that the main contributor to the production of vertical shear in the convection zone is the tilting term of vorticity, $T^{\omega}_1$. True vertical vortices, on the other hand, are mainly produced by baroclinic effects. In the photosphere and chromosphere, both methods suggest magnetic terms to be the main responsible for the generation of vortex flows. Nevertheless, the swirling equation shows that both the magnetic baroclinic and the magnetic tension terms are equally important, while according to vorticity the magnetic tension dominates alone. We can therefore infer that shears in these layers are mainly produced by magnetic tension effects.


\subsection{Analysis of the swirling equation terms}
\begin{figure}
	\centering
	\resizebox{\hsize}{!}{\includegraphics{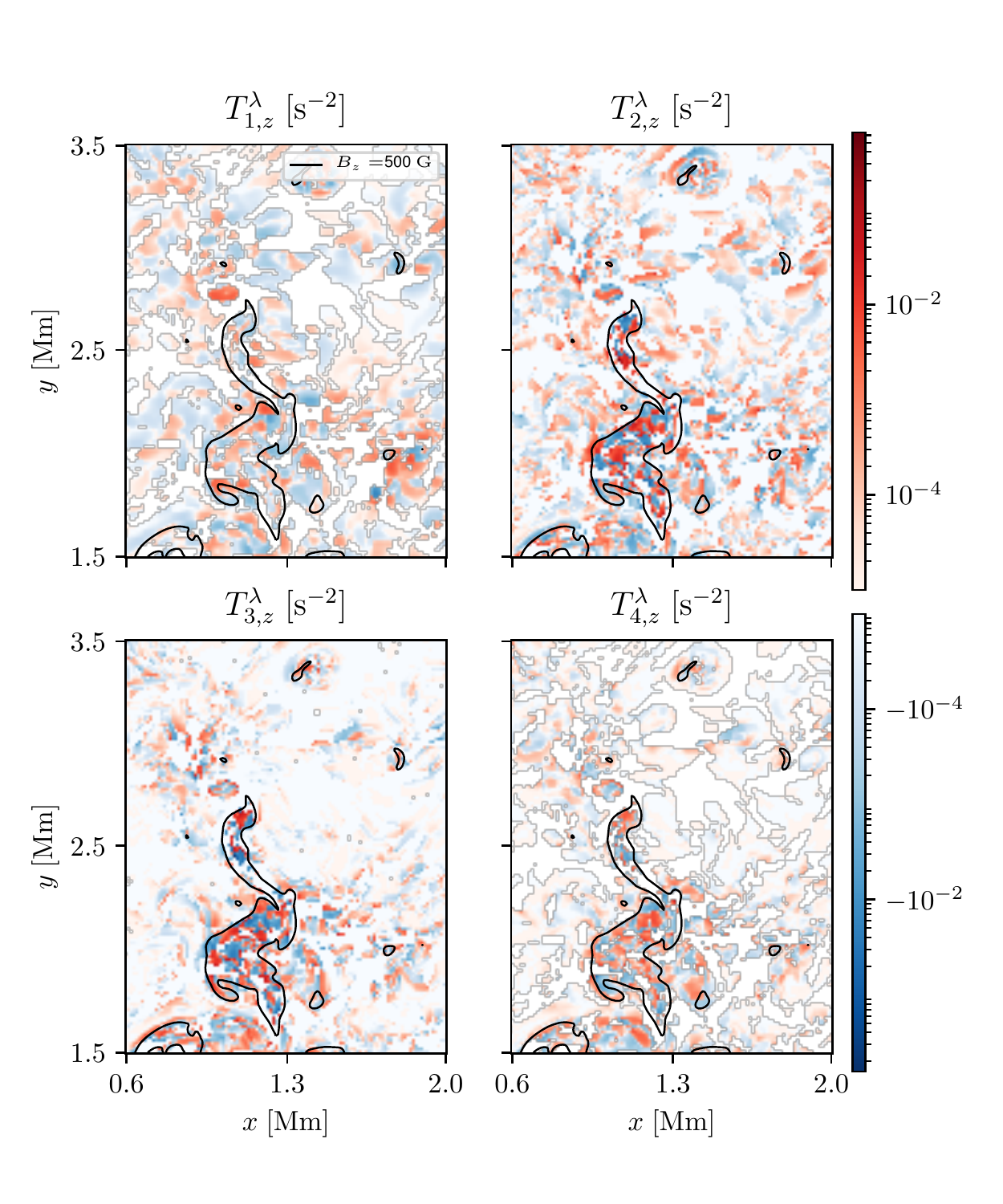}}
	\caption{
	Vertical component of the swirling equation terms in a horizontal section in the convection zone at $z=-1.0\,\mathrm{Mm}$. The section encompasses the photospheric vortex and surroundings of Fig.~\ref{fig:shearflows}. The black contours show the boundaries of strong vertical magnetic flux concentrations with $B_z \ge 500\,\mathrm{G}$.
	}
	\label{fig:ConfrontTerms_25}
\end{figure}

\begin{figure}
	\centering	
	\resizebox{\hsize}{!}{\includegraphics{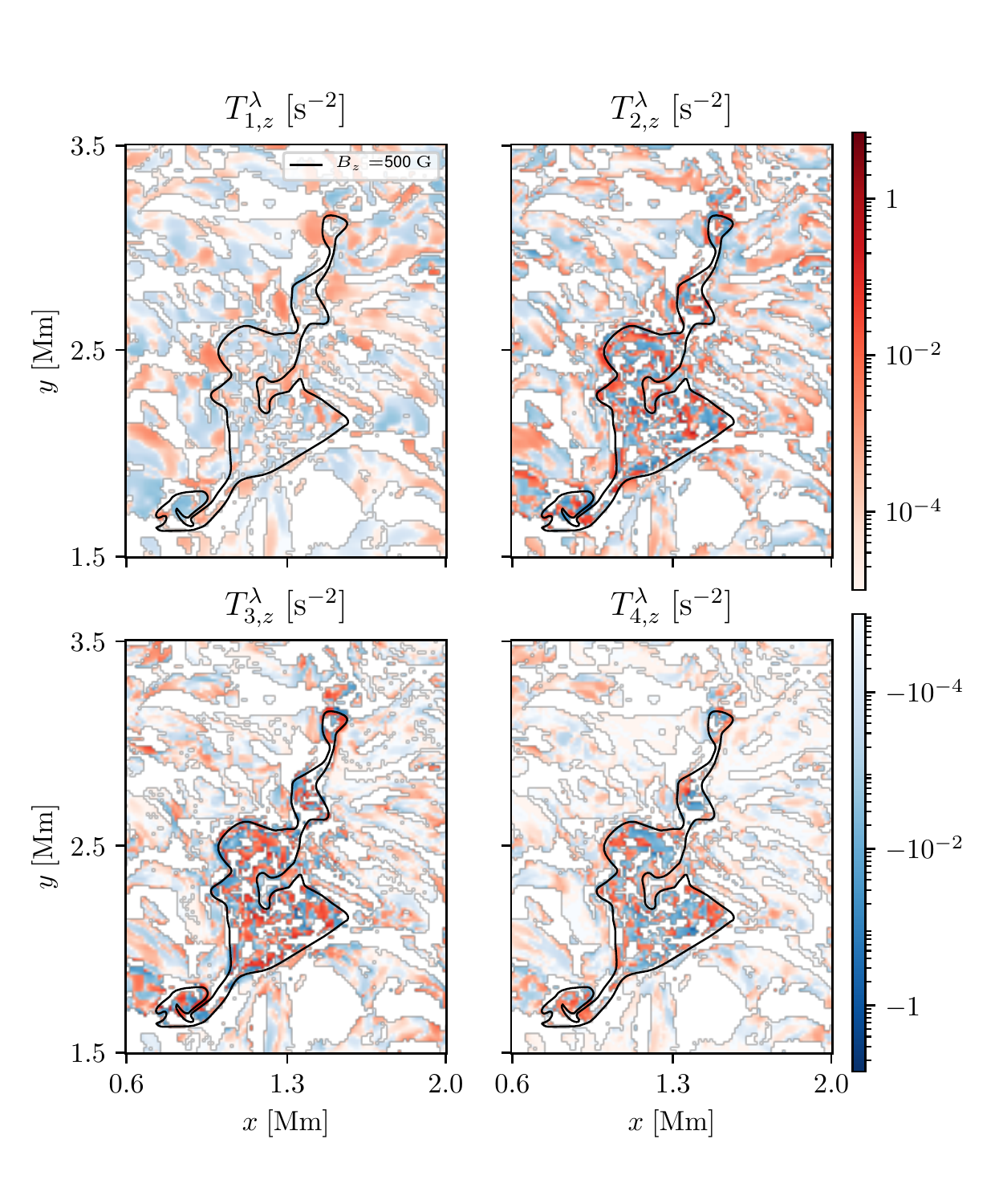}}
	\caption{
	Vertical component of the swirling equation terms in a horizontal section at the optical surface at $z=0\,\mathrm{Mm}$. The section encompasses the photospheric vortex and surroundings of Fig.~\ref{fig:shearflows}. The black contour shows the boundary of strong vertical magnetic flux concentration with $B_z \ge 500\,\mathrm{G}$.
	}
	\label{fig:ConfrontTerms_125}	
\end{figure}

\begin{figure}
	\centering	
	\resizebox{\hsize}{!}{\includegraphics{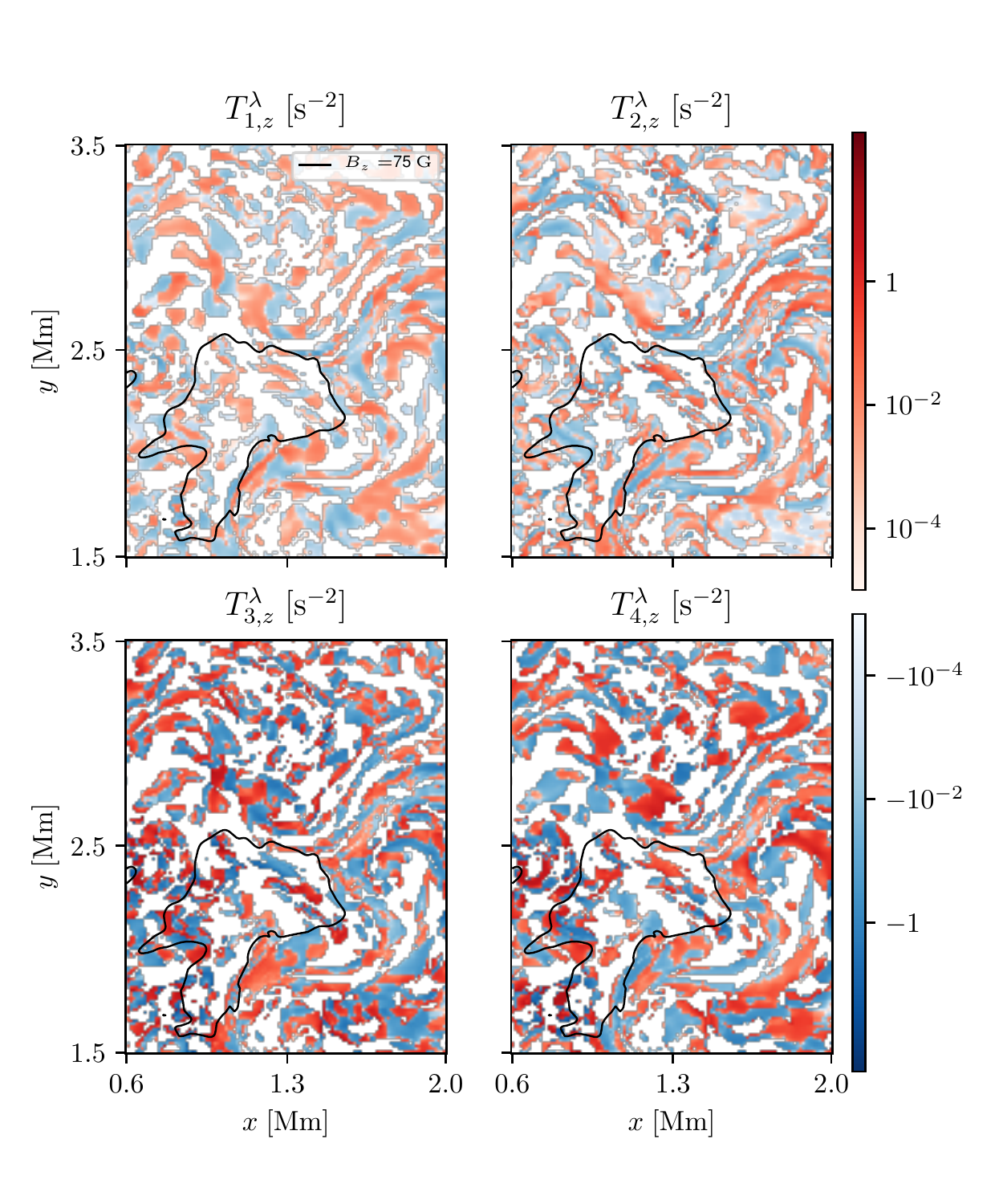}}
	\caption{
	Vertical component of the swirling equation terms in a horizontal section in the chromosphere at $z=1.5\,\mathrm{Mm}$. The section encompasses the photospheric vortex and surroundings of Fig.~\ref{fig:shearflows}. The black contour shows the boundary of strong vertical magnetic flux concentration with $B_z \ge 75\,\mathrm{G}$.
	}
	\label{fig:ConfrontTerms_275}
\end{figure}

We now focus on the swirling equation and see how the different terms locally contribute to the generation of vertical vortices. For this purpose, we  study in more detail the photospheric vortex presented in Fig.\,\ref{fig:shearflows} at three characteristic heights. 
We recall that, like in Sect.\,\ref{subsec:vortvsswirl}, we examine only the vertical component of the 
first term on the right-hand side of the evolution equation for the swirling vector, Eq.\,(\ref{eq:totalswirlingeq}),
where we 
look at the various terms of $\rm{d}\lambda/{\rm d}t$ as given by the swirling equation, Eq.\,(\ref{eq:swirling_eq}).
In what follows, positive values represent the generation of counter-clockwise swirling strength along the $\hat{z}$-axis, while negative ones stand for clockwise perturbations.

Figure\,\ref{fig:ConfrontTerms_25} shows the strength of the various swirling equation terms in a horizontal section in the convection zone of the subdomain under consideration, more precisely, at $z = -1.0\,\mathrm{Mm}$ below the mean optical surface, $\tau_{500} = 1$. It becomes clear that the two terms that show the strongest vertical values are $T^{\lambda}_2$ and $T^{\lambda}_3$, the hydrodynamical and magnetic baroclinic terms, as it was expected from the results of Sect.\,\ref{subsec:vortvsswirl}. The regions which exhibit strong generation of vertical swirling strength by these two terms harbor a strong magnetic field; in fact, the black contours show the boundaries of magnetic flux concentrations with $B_z \ge 500\,\mathrm{G}$. This coexistence can be easily understood regarding the magnetic baroclinicity and the magnetic tension, since their terms are directly related to the magnetic field strength and its variations. Concerning $T^{\lambda}_2$ instead, we infer from this figure that the magnetic flux concentration is generating strong hydrodynamical baroclinicity. A close examination reveals that, inside the boundaries of the magnetic flux concentrations, the structure of $T^{\lambda}_2$ is essentially equivalent to the  magnetic baroclinic one, $T^{\lambda}_3$, but with opposite sign. This picture confirms that the two baroclinic terms are partially counter-balancing each other and that this phenomenon is due to mechanical balance of pressures across the magnetic flux-tube boundary.  

Outside the magnetic flux concentration, the two magnetic terms are weak compared to the hydrodynamical baroclinic one, which contributions are probably due to turbulent motions of the plasma. The stretching term $T^{\lambda}_1$ is generally weaker than the other terms, in particular inside the magnetic flux concentration, which is again in line with what has been previously found in Sect.~\ref{subsec:vortvsswirl}.

The same analysis has been carried out at the mean optical surface ($z=0$), which is shown in Fig.\,\ref{fig:ConfrontTerms_125}. In this layer, the dominant terms are still the baroclinic ones, $T^{\lambda}_2$ and $T^{\lambda}_3$, which are particularly strong inside the magnetic flux concentration. They show again a similar configuration but with opposite sign with respect to each other. At $z=0$, the magnetic tension term starts to be comparable in strength to the baroclinic terms, in particular where the magnetic field is strong. Finally, the stretching term, $T_1^{\lambda}$, is subdominant also in the photosphere. 

The picture changes drastically as we move to the chromosphere, at $z=1.5\,\mathrm{Mm}$ above the optical surface, as shown in  Fig.\,\ref{fig:ConfrontTerms_275}. Here, the swirling strength is principally produced by magnetic processes. In fact, the difference between the hydrodynamical terms, $T^{\lambda}_1$ and $T^{\lambda}_2$, and the magnetic ones, $T^{\lambda}_3$ and $T^{\lambda}_4$, is evident. At this height, in the solar atmosphere, the stretching and hydrodynamical baroclinic terms are comparable, as we can see from Fig.\,\ref{fig:terms_all}. Furthermore, the strength of the magnetic terms is not related to the magnetic field strength any more. In fact, as we can see in Fig.\,\ref{fig:ConfrontTerms_275}, the contributions of $T^{\lambda}_3$ and $T^{\lambda}_4$ to the production of swirling strength are not concentrated where the magnetic field is strong, but are rather homogeneously dispersed over the full cross section.

An interesting peculiarity of the swirling equation terms is their patchy structure, as can be seen in Figs.\,\ref{fig:ConfrontTerms_25}, \ref{fig:ConfrontTerms_125} and \ref{fig:ConfrontTerms_275}: they all share a pattern of sharp and chaotic change of sign from one neighboring small-scale region to another, and this structure appears at all characteristic layers. The size of the patches appears to be smaller in the convection zone ($l \lesssim 50\,\mathrm{km}$) and, in general, in the concentrations of the magnetic field (see Fig.\,\ref{fig:ConfrontTerms_125}). In the chromosphere instead, the patches are on average slightly larger ($l \approx 100\,\mathrm{km}$).

The swirling strength does not exhibit these patches (see, e.g., panel (b) of Fig.\,\ref{fig:shearflows}), which means that the contributions by the terms of the swirling equation generally do not accumulate but they do average to zero in time. This picture may indicate that a large part of these small-scale perturbations are chaotically produced. This situation is reminiscent of waves, which amplitudes are periodically enhanced and reduced by sinusoidal sources. Therefore, these patches could also be signatures of torsional Alfvén waves and Alfvén-type motions in the solar atmosphere. In fact, Figs. \ref{fig:ConfrontTerms_25}, \ref{fig:ConfrontTerms_125} and \ref{fig:ConfrontTerms_275} show the instantaneous generation of both clockwise and counter-clockwise swirling strength at a specific time of the simulation. In the convection zone and low photosphere, the production of a clockwise swirling motion in the plasma, followed by a counter-clockwise one some time after, would induce a torsional wave in the ``frozen-in'' magnetic field. Conversely, Alfvén waves in the chromosphere would induce successive positive and negative values in the swirling strength, produced by the magnetic terms of the
swirling equation.

Torsional Alfvén waves have been observed by \citet{2009Sci...323.1582J} 
in a large bright point group with area $430\,000\,\rm{km^2}$. This supports our hypothesis because the perturbations in the swirling strength that we see are mainly confined within the boundaries of a vertically directed magnetic flux concentration. Moreover, 
\citet{2013ApJ...776L...4S} 
found signatures of propagating Alfvén waves in numerical simulations of the photosphere, while \citet{2015A&A...577A.126M} 
have shown that Alfvén waves can propagate within magnetic flux tubes from the photosphere to the corona. \citet{battaglia2020} found from magneto-hydrodynamic simulation data unidirectional swirls propagating with Alfv\'en speed from the photosphere to the chromosphere, being generated by rotational motions at their magnetic footpoints. From the results presented in this paper we cannot confirm the presence of these waves in our simulations. However, the possibility that the small-scale perturbations we observe, or at least part of them, are signatures of these waves is exciting and deserves a more detailed study.


\section{Conclusion}\label{sec:conclusion}

In this paper, we discussed the importance of ``shear-free'' detection criteria for a proper interpretation of vortices. Vorticity was the only criterion for which an evolution equation was known so far and, therefore, the only possible choice in order to study the dynamics of vortices. Since vorticity can lead to erroneous detections of vortices in turbulent flows, its evolution equation is in principle also not reliable for dynamical investigations. For this reason, we derived an evolution equation for the swirling strength, which is a more reliable choice for vortex identification and for which we provide a detailed description and physical interpretation.

The swirling equation represents a new theoretical result in  (magneto-)hydrodynamics. It can be easily generalized by modifying the underlying momentum equation, for example by adding viscosity terms or by removing the magnetic fields. With this equation, the dynamics of vortices can now be studied in more detail and confidence, since the presence of shear flows does not bias the results.

As a first application of the swirling equation, we investigated which one of its source terms is predominant in numerical simulations of the solar atmosphere. 
We found that, in the convection zone, the hydrodynamic and magnetic baroclinic terms dominate the production of the vertical component of the swirling strength. Concerning vertical vorticity, the tilting term dominates its production.
Higher up, both the production of vertical swirling strength and vorticity is essentially due to magnetic effects. Nevertheless, while magnetic tension is the main factor contributing to the evolution of vorticity, for the swirling strength, both the magnetic baroclinic and the tension term are equally important.

We also showed the vertical component of the swirling equation terms in horizontal sections of a small portion of a simulation model at three characteristic heights. We found that, in the convection zone and in the photosphere, the production of vertical swirling strength is mainly located in strong concentrations of magnetic field. Furthermore, the hydrodynamic and magnetic baroclinic terms often have opposite effects canceling each other. In the chromosphere instead, the magnetic terms alone dominate and produce swirling strength virtually everywhere, in a rather isotropic fashion and quite independently of the location of the magnetic flux concentration in the photosphere. 
These figures also reveal that all the terms share a common small-scale patchy structure, which implies the production of patches of swirling strength of opposite orientation in almost contiguous regions. These swirls could possibly be caused by turbulent motions in the plasma, however, since their strength is increased within magnetic flux concentrations, they could indicate the presence of torsional Alfvén waves.

The swirling strength criterion and its equation are directly applicable to numerical simulations, since the velocity field and the other quantities are known at each time step and spatial point. From observations, the three-dimensional velocity field could be derived by combining Doppler measurements with local correlation or feature tracking. However, a single wavelength measurement would be limited to a certain optical surface. 
The three-dimensional velocity gradient tensor would be derived from observations in multiple spectral lines and wavelength regions, but it would be limited in accuracy and afflicted by noise.
Nevertheless, in a first step, one could limit the application of the swirling strength criterion to a two-dimensional optical surface. In this case, the detection of swirls in that surface would be based on the two-dimensional velocity gradient tensor.

We expect that imaging and spectropolarimetry with the new Daniel K.~Inouye Solar Telescope (DKIST) will reveal details of photospheric and chromospheric swirls of unprecedented high spatial and temporal resolution. Therefore, a deeper understanding of the physics related to vortices in the solar plasma is essential. In this sense, we believe that the swirling equation derived in the present paper is an important step forward to the understanding of the underlying mechanisms and processes of vortical motions in the solar atmosphere.


\begin{acknowledgements} 
This work was supported by the Swiss National Science
Foundation under grant ID 200020\_182094. Special thanks are addressed to A.~Battaglia, A.~Bossart, G.~Janett, P.~Rajaguru, and G.~Vigeesh for enriching comments and discussions. The numerical simulations were carried out by F.~Calvo on Piz Daint at CSCS under project ID s560.
This work has profited from discussions with the team of K.~Tziotziou and E.~Scullion (conveners) ``The Nature and Physics of Vortex Flows in Solar Plasmas'' at the International Space Science Institute (ISSI). O.S.~wishes to acknowledge scientific discussions within the WaLSA team (Waves in the Lower Solar Atmosphere; https://www.WaLSA.team).
The  authors are grateful to the anonymous referee for valuable comments that helped improving the article.
\end{acknowledgements}

\bibliographystyle{aa} 
\bibliography{biblio.bib} 


\begin{appendix}

\section{Physical interpretation of the swirling strength}\label{app:physical_interp}

The swirling strength is a precise criterion for vortex identification, however its convoluted definition causes its physical interpretation to be less straightforward than that of vorticity. This appendix is intended to shed some light on its mathematical definition and physical meaning.

We start from Eq.\,(\ref{eq:U_decomposition}), which shows the explicit diagonalization of the velocity gradient tensor $\mathcal{U}$. As becomes clear further down, whenever the flow is rotating, $\mathcal{U}$ has one real eigenvalue, $\lambda_{\rm r}$, and two complex conjugate ones, $\lambda_+ = \left(\lambda_-\right)^*$. Since $\lambda_+$ and $\lambda_-$ are complex conjugates, so are their relative eigenvectors, 
$\vec{u}_+ =  \left(\vec{u}_-\right)^*$. To prove it, we take the defining relation for the $\vec{u}_+$ eigenvector, $\mathcal{U}\vec{u}_+ = \lambda_+\vec{u}_+$, and we complex conjugate it. We then obtain,
\begin{align*}
  \left( \mathcal{U}\vec{u}_+\right)^* &= \left(\lambda_+\vec{u}_+ \right)^*, \\
  \mathcal{U}\left(\vec{u}_+\right)^* &= \lambda_-\left(\vec{u}_+\right)^*,
\end{align*}
where we use the fact that $\mathcal{U}$ is a real matrix and $\lambda_+ = \left(\lambda_-\right)^*$. The last equation proves our proposition, since that relation is 
equivalent to the defining relation for the $\vec{u}_-$ eigenvector,
$\mathcal{U}\vec{u}_- = \lambda_-\vec{u}_-$.

Yet, in this formulation, both $\Lambda$ and $\mathcal{P}$ appearing in Eq.\,\ref{eq:U_decomposition} are complex matrices, therefore, a physical interpretation is difficult to grasp. Consequently, we would like to transform them into real matrices; an easy way to do it is to use the following transformations, 
\begin{align}
    & \vec{u}_{\rm cr} = \frac{1}{\sqrt{2}}\left(\vec{u}_+ + \vec{u}_-\right)\,,\quad\vec{u}_{\rm ci} = \frac{1}{\sqrt{2}\rm{i}}\left(\vec{u}_+ - \vec{u}_-\right)\,, \nonumber \\
    & \lambda_{\rm cr} = \frac{1}{2}\left( \lambda_+ + \lambda_- \right)\,,\quad\lambda_{\rm ci} = \frac{1}{2\rm{i}}\left( \lambda_+ - \lambda_- \right)\,. \nonumber 
\end{align}
Then, we can prove that the velocity gradient tensor $\mathcal{U}$ can be decomposed in the following form,
\begin{eqnarray}
\mathcal{U} =  \underbrace{\vphantom{\begin{bmatrix}
		\lambda_{\rm r} & 0 &  0\\
		0 & \lambda_{\rm cr}& \lambda_{\rm ci} \\
		0 & \lambda_{\rm ci} & \lambda_{\rm cr}  
		\end{bmatrix}} \left[ \vec{u}_{\rm r}, \vec{u}_{\rm cr}, \vec{u}_{\rm ci}\right] }_{\mathcal{\Tilde{P}}}
	\underbrace{\begin{bmatrix}
\lambda_{\rm r} & 0 &  0\\
		0 & \lambda_{\rm cr}& \lambda_{\rm ci} \\
		0 & -\lambda_{\rm ci} & \lambda_{\rm cr} 
\end{bmatrix}}_{\Tilde{\Lambda}}
\underbrace{\vphantom{\begin{bmatrix}
		\lambda_{\rm r} & 0 &  0\\
		0 & \lambda_{\rm cr}& \lambda_{\rm ci} \\
		0 & \lambda_{\rm ci} & \lambda_{\rm cr}  
		\end{bmatrix}} \left[ \vec{u}_{\rm r}, \vec{u}_{\rm cr}, \vec{u}_{\rm ci} \right]^{-1}}_{\Tilde{\mathcal{P}}^{-1}}\,,  \label{eq:U_realdecomposition}
\end{eqnarray}
where $\Tilde{\mathcal{P}}$, $\Tilde{\Lambda}$, and $\Tilde{\mathcal{P}}^{-1}$ are real matrices. 

We analyze in more detail the new block-diagonal matrix $\Tilde{\Lambda}$, where we can distinguish two blocks: the first one of dimensions $1\times 1$ is associated with the eigenvector $\vec{u}_{\rm r}$. It can be seen as a dilatation operator, since when applied to a scalar it simply acts as a multiplicative constant. The eigenvalue $\lambda_{\rm r}$ can therefore be interpreted as a measure of how the flow is stretched or compressed along the direction given by $\vec{u}_{\rm r}$. 
The $2 \times 2$ block matrix instead can be further decomposed in two pieces,
\begin{equation}
   \begin{bmatrix}  \lambda_{\rm cr}& \lambda_{\rm ci} \\
		           -\lambda_{\rm ci} & \lambda_{\rm cr} \end{bmatrix}
   = \underbrace{\begin{bmatrix}  \lambda_{\rm cr}& 0\\
		                         0 & \lambda_{\rm cr} \end{bmatrix}}_{\Tilde{\mathcal{D}}} + 
	 \underbrace{\begin{bmatrix}  0& \lambda_{\rm ci} \\
		                           -\lambda_{\rm ci} & 0  \end{bmatrix}}_{\Tilde{\mathcal{R}}} \nonumber\,.     
\end{equation}
The $\Tilde{\mathcal{D}}$ matrix can be interpreted as a $\mathrm{2D}$ dilatation operator, for the same reasons we defined the $1\times1$ matrix in this way. The $\Tilde{\mathcal{R}}$  matrix instead represents a infinitesimal rotation operator in the plane spanned by  $\vec{u}_{\rm cr}$ and $\vec{u}_{\rm ci}$, which orientation is clockwise in the direction defined by $\vec{u}_{\rm r}$. Therefore, we can say that $\lambda_{\rm cr}$ measures how stretched is the flow in the rotation plane, while $\lambda_{\rm ci}$ describes the strength of the swirling flow and $\vec{u}_{\rm r}$ the vortex axis: hence the definition of swirling strength. For some visual intuition, the interested reader can refer to \citet{doi:10.2514/6.1999-3288}. 

The mathematical description here presented hides some arbitrariness in what concerns the orientation (clockwise or counter-clockwise) of the swirl. In fact, eigenvectors are, in general, defined up to a multiplicative constant, which results in a arbitrary definition of their norm and orientation. This implies that, mathematically, we can multiply the eigenvector $\vec{u}_{\rm r}$ by $-1$ and it would still be a correct solution of the eigenanalysis. However, physically, this would imply the inversion of the vortex orientation.   

To be able to select the correct orientation of the eigenvector $\vec{u}_{\rm r}$, one has to check the handedness of the new basis. In fact, the matrix $\Tilde{\mathcal{R}}$ describes a clockwise rotation only if the $\mathrm{3D}$ basis in which it is defined is right-handed. The basis in question is not the standard one for a three-dimensional space, but the one composed by the real vectors $\vec{u}_{\rm r}$, $\vec{u}_{\rm ci}$, and $\vec{u}_{\rm cr}$ (in this order). Indeed, when we decompose the velocity gradient tensor $\mathcal{U}$ as in Eq.\,\ref{eq:U_realdecomposition}, we are essentially performing a change of basis. The new basis, embodied in the matrix $\Tilde{\mathcal{P}}$, is proper to the flow. Therefore, to find out if the orientation of the eigenvector is correct, one has to check the orientation of the basis itself. One easy way to do it, is to compute the determinant of the matrix formed by the three basis vectors, that is, $\Tilde{\mathcal{P}}$: if $\rm{Det}(\Tilde{\mathcal{P}}) > 0$ the basis is right-handed, while if $\rm{Det}(\Tilde{\mathcal{P}}) < 0$ it is the opposite. 
In this way, one can fix the orientation of the real eigenvector and ensure that the swirling vector, $\vec{\lambda} \equiv \lambda \vec{u}_{\rm r}$, correctly describes the vortex flow.


\section{Commutator with diagonal matrices}\label{app:determinant_proof}
We prove that the commutator between a diagonal and a generic matrix results in a hollow matrix. In order to do that, here we consider the commutator between two matrices $\mathcal{C}$ and $\mathcal{D}$. If $\mathcal{D}$ is diagonal, its terms can be expressed as $\mathcal{D}_{ij} \equiv d^{(i)}\delta_{ij}$, where $\delta_{ij}$ is the Kronecker's delta and $d^{(i)}$ are the diagonal terms,  
\begin{eqnarray}
    \left[\mathcal{C},\mathcal{D}\right]_{ij} &=& \sum_k\left(\mathcal{C}_{ik}\mathcal{D}_{kj} - \mathcal{D}_{ik}\mathcal{C}_{kj}\right) \,,\nonumber\\
    &=& \sum_k \left(  \mathcal{C}_{ik}d^{(k)}\delta_{kj} - d^{(i)}\delta_{ik}\mathcal{C}_{kj} \right)\,,\nonumber\\
    &=& (d^{(j)}-d^{(i)})\mathcal{C}_{ij} \,, \label{eq:commutator_proof}
\end{eqnarray}
which implies that $\left[\mathcal{C}, \mathcal{D} \right]_{ij} = 0 \text{ for } i=j$.


\section{Alternative derivation of the swirling equation}\label{app:alternative_deriv}

In Sect.\,\ref{sec:swirlingequation} we selected the $(2,2)$ component of Eq.\,(\ref{eq:Lambda_eq}) in order to derive the swirling equation. However, we also could have chosen the $(3,3)$ component. Here, we prove that the two approaches result in the same equation. 

The $(3,3)$ component of Eq.\,(\ref{eq:Lambda_eq}) yields the dynamical equation of the $\lambda_-$ eigenvalue, 
\begin{align}
    \frac{\rm d}{{\rm d}t}\lambda_- &= - \lambda_-^2 + \left(\mathcal{P}^{-1}\mathcal{M}\mathcal{P}\right)_{33}\,, \nonumber \\
    \frac{\rm d}{{\rm d}t}\left(\lambda_{\rm cr} - \rm{i} \lambda_{\rm ci}\right) &= -\left( \lambda_{\rm cr} - \rm{i} \lambda_{\rm ci}\right)^2 + \left(\mathcal{P}^{-1}\mathcal{M}\mathcal{P}\right)_{33}\,\nonumber. 
\end{align}
Therefore, by taking the imaginary part, we get,
\begin{align}
    \frac{\rm d}{{\rm d}t}\lambda =& -2\lambda\lambda_{\rm cr} - 2 \rm{Im}\left( \mathcal{P}^{-1}\mathcal{M}\mathcal{P} \right)_{33} \,, \nonumber
\end{align}
which is equivalent to Eq.\,(\ref{eq:swirling_eq}) but for the last term. Therefore, to prove that the two approaches are equivalent we need to check that,
\begin{equation}
   - \rm{Im}\left( \mathcal{P}^{-1}\mathcal{M}\mathcal{P} \right)_{33} = \rm{Im}\left( \mathcal{P}^{-1}\mathcal{M}\mathcal{P} \right)_{22}\,. \label{eq:proof}
\end{equation}

This can be done by considering the properties of the eigenvectors which compose the matrix $\mathcal{P}$. In Appendix\,\ref{app:physical_interp} we have shown that $\vec{u}_+ = (\vec{u}_-)^*$, then one can prove that a similar relation holds also for the vectors composing $\mathcal{P}^{-1}$. In fact, if we describe $\mathcal{P}^{-1}$ with horizontal vectors, 
\begin{equation}
[\vec{u}_{\rm r}, \vec{u}_{\rm +}, \vec{u}_{\rm -}]^{-1} \equiv \begin{bmatrix} \vec{t}^{\rm \,T}_{\rm r} \\ \vec{t}^{\rm \,T}_{\rm +} \\ \vec{t}^{\rm \,T}_{\rm -} \end{bmatrix}\,, \nonumber
\end{equation}
the reader can check that $\vec{t}_{\rm +} = (\vec{t}_{\rm -})^*$. 

Finally, using this decomposition, we can directly compute the two terms which we want to prove to be equivalent,
\begin{equation}
\mathcal{P}^{-1}\mathcal{M}\mathcal{P} = \begin{bmatrix} \vec{t}^{\rm \,T}_{\rm r} \\ \vec{t}^{\rm \,T}_{\rm +} \\ \vec{t}^{\rm \,T}_{\rm -} \end{bmatrix} \begin{bmatrix} 
                \vphantom{\vec{t}^{\rm \,T}_{\rm r}} m_{11} & m_{12} & m_{13} \\
                \vphantom{\vec{t}^{\rm \,T}_{\rm r}} m_{21} & m_{22} & m_{23} \\
                \vphantom{\vec{t}^{\rm \,T}_{\rm r}} m_{31} & m_{32} & m_{33} \\
            \end{bmatrix}
[\vec{u}_{\rm r}, \vec{u}_{\rm +}, \vec{u}_{\rm -}]\,. \nonumber            
\end{equation}
By carrying out the computation for both $(2,2)$ and $(3,3)$ components, one can check that Eq.\,(\ref{eq:proof}) holds true and that, therefore, the two approaches to the derivation of the swirling equation are equivalent.


\end{appendix}
\end{document}